\documentclass[a4paper]{article}
\usepackage{latexsym}
\begin {document}

\title {\bf A possible Reinterpretation of Einstein's Equations }
\author{A.~Bouda\footnote{Electronic address:
{\tt bouda\_a@yahoo.fr}}
\ and A.~Belabbas \footnote{Electronic address:
{\tt belabbas.moumene@gmail.com }}\\
Laboratoire de Physique Th\'eorique, Universit\'e de B\'eja\"\i a,\\
Route Targa Ouazemour, 06000 B\'eja\"\i a, Algeria\\}

\date{\today}

\maketitle

\begin{abstract}
\noindent
In this paper, we first review Huei's formulation in which it is shown
that the linearized Einstein equations can be written in the same form
as the Maxwell equations. We eliminate some imperfections like the
scalar potential which is ill linked to the electric-type field,
the Lorentz-type force which is obtained with a time independence
restriction and the undesired factor 4 which appears in the
magnetic-type part. Second, from these results and in the light of a
recent work by C.C. Barros, we propose an extension of the equivalence
principle and we suggest a new interpretation for Einstein's equations
by showing that the electromagnetic Maxwell equations can be
derived from a new version of Einstein's ones.
\end{abstract}

\vskip\baselineskip



\newpage

\section{Introduction}

There are several attempts to describe gravitation and
electromagnetism in a unified field theory. The electromagnetic
field was sometimes presented as the nonsymmetric part of the
metric \cite{Einstein1,Einstein2}, sometimes as some additional
components of the metric in five dimensional space-time
\cite{OW}. In this paper, we suggest that the electromagnetic field is
contained in the four dimensional Einstein equations themselves.
Our approach is based on two things.

The first is Huei's formulation \cite{Huei} in which it is shown
that the linearized Einstein equations can be written in the same form
as the Maxwell equations. In this formulation, developed also by Wald
\cite{Wald}, there are however some imperfections to point out:
\begin{enumerate}

\item The usual relation between the potential and the
electric-type field as known in electromagnetism is reproduced only
in the harmonic gauge.

\item The geodesic equation predicts a Lorentz-type force with an undesired
factor 4 in the magnetic-type part compared to the usual electromagnetic Lorentz
force.

\item The Lorentz-type force is obtained only in the case where the fields are
independent on time.

\end{enumerate}
We would like to add that Carroll \cite{Carroll} has redefined the fields in
such a way as to eliminate the undesired factor 4. However, Maxwell-type equations
are not satisfied in his formulation. In what follows, with the use of a subtle
gauge, we will review the linearized gravity so as to get to a strong similarity
with electromagnetism.

The second thing is the recent work by C.C.Barros in which he suggested
to study the effect of the metric in the subatomic systems instead of
trying to quantize gravity
\cite{Barros1,Barros2,Barros3}. He assumed that interactions, even non
gravitational ones, affect the space-time structure. In the context of
Schwarzschild's solution, by introducing Coulomb's potential instead of
the gravitational one into the metric, he derived  the
spectrum of hydrogen atom. Although the author considered a subatomic
system, however the use of Schwarzschild's solution implies that he
implicitly assumed the existence of versions of Einstein's equations
and geodesic equations for the Coulomb interaction.
What is even more surprising is that his results for hydrogen atom
spectrum, confirmed in more detail in \cite{Belabbas}, are extremely
close to those predicted by Dirac's equation. These intriguing results
encourage us to wonder about the extension of the principle of equivalence
(PE).

The Barros results and the extraordinary similarity between gravitation and
electromagnetism resulting from the revised version of linearized gravity
presented here suggest that the electromagnetic Maxwell equations can be
followed from a new version of Einstein's equations in the linear
approximation within the framework of the electromagnetic interaction.
We will see that the higher order terms of this new version of Einstein's
equations are negligible in the usual application domains of the
electromagnetism.

The paper is organized as follows. In section 2, we will review the
linearized gravity. In section 3, we will present several arguments to show
that the current version of the PE is restrictive and suggest how it can
be extended to include other interactions. We consolidate our arguments
by showing in the context of the electromagnetic interaction that
Maxwell's equations are contained as a first order approximation
in Einstein's equations. Section 4 is devoted to conclusion.

\section{Linearized gravity revisited}

Our goal here is to review the linearized gravity so as to
avoid the imperfection cited above and then to get to a strong similarity
with electromagnetism. For this purpose, let us decompose the metric into
the Minkowski one plus a perturbation
\begin {equation}
g_{\mu\nu}=\eta_{\mu\nu}+ h_{\mu\nu},
\end {equation}
where $\eta_{\mu\nu}=(+1,-1,-1,-1)$ and $h_{\mu\nu}\ll 1$. To first order,
setting
\begin {equation}
g^{\mu\nu}=\eta^{\mu\nu}- h^{\mu\nu}
\end {equation}
and taking into account Eq. (1), it is easy to check that indices of
$h^{\mu\nu}$ can be raised and lowered by using $\eta_{\mu\nu}$ and
$\eta^{\mu\nu}$. In this approximation, the Christoffel symbols,
\begin {equation}
\Gamma^{\lambda}_{\mu\nu}={1 \over 2}g^{\lambda\rho}
        \left[
        \partial_{\mu}g_{\rho\nu} + \partial_{\nu}g_{\mu\rho}
                 - \partial_{\rho}g_{\mu\nu}
        \right],
\end {equation}
and the curvature tensor,
\begin {equation}
R^{\lambda}_{\ \ \mu\nu\rho}= \partial_{\rho} \Gamma^{\lambda}_{\mu\nu}
                            - \partial_{\nu} \Gamma^{\lambda}_{\mu\rho}
                + \Gamma^{\sigma}_{\mu\nu}\Gamma^{\lambda}_{\rho\sigma}
                - \Gamma^{\sigma}_{\mu\rho}\Gamma^{\lambda}_{\nu\sigma},
\end {equation}
take the forms
\begin {equation}
\Gamma^{\lambda}_{\mu\nu}={1 \over 2}\eta^{\lambda\rho}
        \left[
        \partial_{\mu}h_{\rho\nu} + \partial_{\nu}h_{\mu\rho}
                 - \partial_{\rho}h_{\mu\nu}
        \right]
\end {equation}
and
\begin {equation}
R_{\lambda\mu\nu\rho}= {1 \over 2} \left[ \partial_{\rho}\partial_{\mu} h_{\lambda\nu}
                                        + \partial_{\nu}\partial_{\lambda} h_{\mu\rho}
                                        - \partial_{\lambda}\partial_{\rho} h_{\mu\nu}
                                        - \partial_{\mu}\partial_{\nu} h_{\lambda\rho} \right].
\end {equation}
The infinitesimal coordinate transformation $x^{\mu}\rightarrow x'^{\mu}=x^{\mu}+\xi^{\mu}$
induces on the perturbation of the metric a transformation
$\delta h_{\mu\nu}=-\partial_{\nu}\xi^{\lambda}\eta_{\mu\lambda}
                   -\partial_{\mu}\xi^{\lambda}\eta_{\nu\lambda} $
which leaves invariant expression (6) of the curvature tensor. This invariance allows
us to fix a gauge
\begin {equation}
g^{\mu\nu}\Gamma^{\rho}_{\mu\nu}=0,
\end {equation}
known as the harmonic gauge \cite{Wald,Carroll,Weinberg}. In the linear approximation,
this condition takes the following form
\begin {equation}
\partial_\mu h^{\mu}_{\nu} - {1 \over 2} \partial_\nu h =0,
\end {equation}
where $h\equiv h^{\mu}_{\ \mu}$ is the trace of the perturbation.
Huei defined the electric-type field and the potential vector as
\begin {equation}
E_{g}^{i}={c^{2} \over 4} \left[\partial^{i}\hat{h}^{00}+\partial_{j}\hat{h}^{ij} \right]
\end {equation}
and
\begin {equation}
A_{g}^{i}={c \over 4} \hat{h}^{0i} ,
\end {equation}
where $\hat{h}^{\mu\nu}=h^{\mu\nu}-h \eta^{\mu\nu}/2$, $c$ is the velocity of light and
$i,j,...=1,2,3$. It is assumed that the Einstein convention for repeated indexes works
also for $i$ and $j$. In this formulation, although expression (9) satisfies
Maxwell-type equations, however it does not allow to reproduce the expected
relation
\begin {equation}
\overrightarrow{E}_{g}=-\overrightarrow{\nabla}\phi-{\partial \overrightarrow{A}_{g} \over \partial t} ,
\end {equation}
where $\phi$ is the newtonian potential. Nevertheless, if we use the gauge condition (8),
expression (9) turns out to be
\begin {equation}
E_{g}^{i}={c^{2} \over 4} \partial^{i}\hat{h}^{00}-{c^{2} \over 4}\partial_{0}\hat{h}^{0i} ,
\end {equation}
which is compatible with (10) and (11) if we set $\phi= c^{2}\hat{h}_{00}/4$.
This feature is unsatisfactory since relation (11) should be valid in any gauge.
We also notice that in the Lorentz-type force,
\begin {equation}
{d^2 \overrightarrow{r} \over dt^2}
             =   \overrightarrow{E}_{g} + 4 \overrightarrow{v} \times\overrightarrow{B}_{g} ,
\end {equation}
obtained from the geodesic equation, there is in the magnetic-type part an undesired
factor 4 compared to the electromagnetic case. In addition, this equation is
obtained only if the fields do not depend on time.

In what follows, we will remedy these weakness by defining the scalar and
vector potentials as
\begin {eqnarray}
A_{g}^{0}  & = &   {  c \over 2 } h^{00},  \\
A_{g}^{i}  & = &   c  h^{0i}
\end {eqnarray}
and introducing the tensor
\begin {equation}
F_{g}^{\mu\nu}=\partial^{\mu} A_{g}^{\nu}-\partial^{\nu} A_{g}^{\mu} ,
\end {equation}
as in electromagnetism. We notice that definition (14) of the scalar
potential is compatible with the well-known expression obtained when the
geodesic equation is used in the newtonian approximation.
Contrary to Huei's formulation, by defining $E_{g}^{i}= -c F_{g}^{0i}$ as in
electromagnetism, relation (11) is automatically satisfied without using
any gauge. It is also the case for
\begin {equation}
\overrightarrow{B}_{g}=\overrightarrow{\nabla}\times \overrightarrow{A}_{g} ,
\end {equation}
where $B_{g}^{i}= - \epsilon^{ijk} (F_{g})_{jk}/2$ is the magnetic-type field and
$\epsilon^{ijk}$ is the usual Levi-Civita antisymmetric tensor
$\left( \epsilon^{123}=+1 \right)$. Concerning the first group of the Maxwell-type
equations,
\begin {eqnarray}
\partial^\lambda F_{g}^{\mu\nu}+ \partial^\nu F_{g}^{\lambda \mu} +
                             \partial^\mu F_{g}^{\nu \lambda}  = 0,
\end {eqnarray}
with the use of definition (16), it is automatically satisfied.
With regard to the second group, let us consider the Einstein tensor
\begin {equation}
G_{\mu\nu} \equiv R_{\mu\nu}- {1\over 2}g_{\mu\nu} R
\end {equation}
and use the following definition for the Ricci tensor
\begin {equation}
R_{\mu\nu}=g^{\lambda\rho}R_{\lambda\mu\rho\nu}.
\end {equation}
In the linear approximation, by using (1), (2), (6) and (20), expression
(19) turns out to be
\begin {eqnarray}
G_{\mu\nu}= {1\over 2} \left[
            \partial_{\mu} \partial_{\nu} h +
             \Box h_{\mu\nu}-
            \eta_{\mu\nu}\Box h  \right.
            \hskip20mm&& \nonumber \\
      \left. - \partial_\lambda \partial_\mu h^\lambda_\nu -
               \partial_\lambda \partial_\nu h^\lambda_\mu +
               \eta_{\mu\nu}\partial_\rho \partial_\lambda h^{\rho\lambda}
                \right],
\end {eqnarray}
where $ \Box = \partial^\lambda \partial_\lambda$ is the d'Alembertian.
Carroll \cite{Carroll} showed that the trace $h^{i}_{\ i}$ is not a
propagating degree of freedom. Then, instead of the time
component of (8), let us impose an alternative condition
\begin {equation}
h^{i}_{\ i}=0
\end {equation}
and keep in the gauge condition (8) only the three space components
\begin {equation}
\partial_\mu h^{\mu}_{i} - {1 \over 2} \partial_{i} h =0.
\end {equation}
From relation (22), we observe that $h=h_{00}$ and expressions (14) and
(15) can be written in a unified form
\begin {equation}
A_{g}^{\mu}=  c \left( h^{0\mu}-{ 1 \over 2 } \eta^{0\mu} h \right).
\end {equation}
With the use of (22), we can deduce from (21) that
\begin {equation}
G_{00}= {1\over 2} \partial_i \partial_j h^{ij}
\end {equation}
and
\begin {eqnarray}
G_{0i}= {1\over 2} \left[ \partial_{j} \partial^{j} h_{0i} -
                          \partial_{0} \partial^{j} h_{ij} -
                          \partial_{i} \partial^{j} h_{0j}
                   \right].
\end {eqnarray}
Relations (22) and (23) allow to write (25) and (26) as follows
\begin {equation}
G^{00}= {1\over 2} \partial_{i} \left[ {1\over 2}
                                \partial^{i} h^{00} - \partial^{0} h^{0i}
                                \right]
\end {equation}
and
\begin {equation}
G^{0i}= {1\over 2} \left[ \partial_{j} \left( \partial^{j} h^{0i} -
                                              \partial^{i} h^{0j} \right) +
                          \partial_{0} \left( \partial^{0} h^{0i} -
                                              {1\over 2} \partial^{i}  h^{00} \right)
                   \right].
\end {equation}
Taking into account (16) and (24), relations (27) and (28) become
\begin {equation}
G^{00}  =  {1 \over 2 c } \partial_{i} F_{g}^{i0}
        =  {1 \over 2 c } \partial_{\mu} F_{g}^{\mu 0}
\end {equation}
and
\begin {equation}
G^{0i} = {1 \over 2 c } \partial_{\mu} F_{g}^{\mu i}.
\end {equation}
Finally, we have
\begin {equation}
G^{0\nu} = {1 \over 2 c } \partial_{\mu} F_{g}^{\mu \nu}.
\end {equation}
It is clear that in the vacuum, Einstein's equations, $G^{0\nu}=0$,
reduce to the Maxwell-type equations $\partial_{\mu} F_{g}^{\mu \nu}=0$.
We notice that the time component of (8), which has not been used,
represents the Lorentz-type gauge $\partial_{\mu}A^{\mu}_{g}=0$.
The last point concerns the geodesic equation
\begin {equation}
{d^2x^\lambda \over d\tau^2} +
    \Gamma^\lambda_{\mu\nu}{dx^\mu \over d\tau}{dx^\nu \over d\tau} = 0,
\end {equation}
which allows us to write in the linear approximation
\begin {eqnarray}
{d^2 x^{i} \over d\tau^2}  =  c^{2} \left[
                  {1 \over 2}\left( {dt \over d\tau}\right)^{2}\partial^{i} h^{00}
                  - {1 \over c} \left( {dt \over d\tau}\right)^{2}
                  {\partial h^{0i} \over \partial t}
                  \right.
                  \hskip30mm&& \nonumber \\
                  - {u_{j} \over c} {dt \over d\tau }
                  \left(\partial^{j} h^{0i}-\partial^{i} h^{0j} \right)
                   \left. - {u_{j} \over c^{2}}  {dt \over d\tau}
                  {\partial h^{ij} \over \partial t} -
                  {u_{j}u_{k} \over c^{2}}
                  \left(\partial^{j} h^{ik}- {1 \over 2}\partial^{i} h^{jk}\right)
                   \right],
\end {eqnarray}
$\tau$ being the proper time and $u_{\nu}$ a component of the four-velocity.
For lower velocities, we have $dt/d\tau \rightarrow 1$ and
$u^{i} \rightarrow v^{i}=dx^{i}/dt$. Then, if we neglect the terms proportional
to $1/c^{2}$ in (33) and use (24), we get to
\begin {equation}
{d^2 x^{i} \over dt^2}
             =    \left(
                         c \ \partial^{i}A_{g}^{0}-  {\partial A_{g}^{i} \over \partial t}
                             \right)
                          - v_{j} \left(
                              \partial^{j} A_{g}^{i} - \partial^{i} A_{g}^{j}
                              \right) ,
\end {equation}
from which we deduce that
\begin {equation}
{d^2 \overrightarrow{r} \over dt^2}
             =   \overrightarrow{E}_{g} + \overrightarrow{v} \times\overrightarrow{B}_{g} .
\end {equation}
We emphasize that this result is obtained without imposing the time
independence restriction for the fields as the case in Refs. \cite{Huei,Wald}.
Furthermore, the undesired factor 4 in the magnetic-type part disappears.

\section{ Maxwell's equations from the Einstein ones }

This extraordinary similarity between gravitation and electromagnetism
obtained in the last section prompt us to wonder about the extension
of the PE. In fact, the PE states that at every space-time point
in an arbitrary gravitational field it is possible to choose a
"locally inertial coordinate system" such that, within a sufficiently
small region of the point in question, the laws of nature take the
same form as in unaccelerated Cartesian coordinate systems
\cite{Weinberg}. Let us provide further arguments in favor of an
extension of the PE.

First, we think that there is some vagueness about the expression
"sufficiently small region". In fact, Cartesian coordinate system
implies vanishing values for the Christoffel symbols
$\Gamma^{\lambda}_{\mu\nu}$.
From a geometrical point of view, it is well-known that at any
point it is possible to impose a vanishing value for the
$\Gamma^{\lambda}_{\mu\nu}$ by a suitable coordinate system. The
non-vanishing value of the curvature tensor, expression (4),
is then ensured by the derivatives of $\Gamma^{\lambda}_{\mu\nu}$.
However, if we substitute the point by a small region, the
vanishing values of $\Gamma^{\lambda}_{\mu\nu}$ will impose also
vanishing values for their derivatives in this region and then
for the curvature tensor, which must keep this value in any
coordinate system. This conclusion is absurd since we assumed
from the start the presence of the gravitational field. Finally, to
be precise, the PE works just at a point and not at a "small region"
providing then the possibility of canceling any interaction, even
the non gravitational one, by a suitable coordinate system.

Second, even if we consider the equality between the inertial mass
and the gravitational one is perfectly exact, two bodies A and B
do not fall to Earth with the same acceleration in the same frame.
In fact, the same acceleration can be measured for the two bodies in
two different frames associated with the mass centers of each body
and the Earth. It is only in the case of the same mass for A and B or
in the case where each body mass is negligible with respect
to the Earth one that the two mass centers coincide. This means
that the property from which all bodies fall with the same
acceleration is just an approximation which works in particular
circumstances.

The third argument is the Einstein unified theory
\cite{Einstein1,Einstein2}. Although he presented the PE as the
cornerstone upon which general relativity is based, he attempted to
extend his theory in order to include the electromagnetic field
without concerning himself with the PE. That's one way of admitting
that the PE can be extended to other interactions.

The last and main argument is the Barros \cite{Barros1,Barros2,Barros3}
result concerning the hydrogen atom. He reproduced its spectrum
by introducing Coulomb's potential into the metric. Because of
the spherical symmetry, he used the Schwarzschild solution
\begin {equation}
ds^2 =\xi_{e} c^{2} dt^{2} -
      r^2(d\theta^2+\sin^2 \theta d\phi^2) - \xi_{e}^{-1} dr^2,
\end {equation}
where $\xi_{e}(r)$ is determined by Coulomb's interaction
$V(r)=-K e^{2} /r$, and he showed that for weak potential
\begin {equation}
\xi_{e}(r)=1-{2 K \over c^2 }{ e \over r}{e \over m_e },
\end {equation}
$m_e$ being the electron mass. Observe that in the gravitational
case we have
\begin {equation}
\xi_{g}(r)=1-{2G \over c^{2}} { M\over  r}{m_{g}\over m_{i}},
\end {equation}
in which the ratio between the gravitational and the inertial
masses is given by $m_{g}/ m_{i}=1$.
The author tacitly assumed that the above expression of $ds^2$
follows from the Einstein equation version of electromagnetic field.
This suggests that a particle motion under the electric force
can be described by a geodesic equation. In fact, let us consider
a charge $q$ of mass $m$ under the action of a fixed charge $Q$.
In the weak field case and lower velocities $v \ll c$,
we can deduce from (32) that
\begin {equation}
{d^2 \overrightarrow{r} \over dt^2} =
            - { c^2 \over 2} \overrightarrow{\nabla} h_{00}.
\end {equation}
If we set
\begin {equation}
h_{00} = { 2K \over c^2} { Q q \over m} { 1 \over r},
\end {equation}
we reproduce the motion equation in agreement with the second
Newton law. The linear approximation is used here and will be
justified below.

All these arguments suggest us two things. The first concerns the possible
extension of the PE in such away as to remove the action of any fundamental
interaction by a suitable coordinate system. We will come back to this
question in the last section. The second suggestion is that Maxwell's
equations are contained as a first order approximation in a new version of
Einstein's ones corresponding to the electromagnetic context. In fact,
as in gravity, let us define in presence of a test particle of mass
$m$ and charge $q$ the four-vector potential as
\begin {equation}
 A^{\mu}= { m c \over q } \left( h^{0\mu}-{ 1 \over 2 } \eta^{0\mu} h \right),
\end {equation}
where $h^{\mu\nu}$ satisfy conditions (22) and (23) as in above.
The well-known definition of the electromagnetic tensor,
$F^{\mu\nu}=\partial^{\mu} A^{\nu}-\partial^{\nu} A^{\mu}$ , guarantees
automatically the first group of Maxwell's equations,
\begin {eqnarray}
\partial_\lambda F_{\mu\nu}+ \partial_\nu F_{\lambda \mu} +
                             \partial_\mu F_{\nu \lambda}  = 0,
\end {eqnarray}
and the usual relations between potentials and fields. With regard to the
second group, by following the same procedure as in section 2
and using (41) instead of (24) in (21), we get to
\begin {equation}
G^{0\nu} = {q \over 2 m c } \partial_{\mu} F^{\mu \nu}.
\end {equation}
It is clear that in the vacuum, Einstein's equations, $G^{0\nu}=0$,
lead to Maxwell's equations $\partial_{\mu} F^{\mu \nu}=0$.

In presence of charges, as in gravity, let us write Einstein's equations
in the following manner
\begin {equation}
G_{\mu\nu}= \chi_{e} T_{\mu\nu},
\end {equation}
where $T_{\mu\nu}$ is a tensor describing the presence of charges and
currents. The constant $\chi_{e}$ must be determined by requiring that
(44) reproduces Maxwell's equations
\begin {equation}
\partial_{\mu} F^{\mu \nu} = \mu_{0} j^{\nu},
\end {equation}
where $(j^{\nu})=(c \rho, j^{i})$ is the
four-current vector. Using (43) and (45) in (44), we obtain
\begin {equation}
\chi_{e} T^{0\nu} = {q \over 2 m c } \mu_{0} j^{\nu}.
\end {equation}
By defining in the linear approximation
\begin {equation}
T^{0\nu} = {q \over  m  }  j^{\nu},
\end {equation}
we deduce that
\begin {equation}
\chi_{e} = {1 \over 2 c^{3} \varepsilon_{0} }.
\end {equation}
It is clear that Maxwell's equations are followed from a new
version of Einstein's equations in the linear approximation.
Let us show that the higher order terms are negligible.
From (41), the electric and magnetic fields are
given by
\begin {equation}
E^{i}= -c F^{0i}
     = - {m c^{2} \over q} \left( \partial^{0}h^{0i}-
                               {1 \over 2} \partial^{i}h^{00}\right)
\end {equation}
and
\begin {equation}
B^{i}= - {1 \over 2} \epsilon^{ijk} F_{jk}
     = - {m c \over 2q} \epsilon^{ijk} \left( \partial_{j}h_{0k}-
                                            \partial_{k}h_{0j} \right) .
\end {equation}
In order to avoid the quantum effects, we do not choose a subatomic
system as a test particle. In the domain of high voltage, let us consider
a dust particle with a diameter of $10^{-6} \ \textrm{m}$. Its mass is in
the order of $m \approx 10^{-14} \ \textrm{kg}$. This particle can be in
an electrostatic precipitator where can reign a strong electric field $E$ or in an
air gap of a rotating electrical machine where can reign a strong magnetic field
$B$. For such a particle, it is well-known \cite{Parker} that its saturation
electric charge is in the order of $q \approx 10^{-18} \ \textrm{C}$. Even for
the field values $E \approx 10^{8} \ \textrm{V} .\textrm{m}^{-1}$ and
$B \approx 10 \ \textrm{T}$ which are rarely reached, it follows that the quantities
$\partial_{\nu}h^{0\mu}$ appearing in (49) and (50) take very small values, in
the order of $qE^{i}/mc^{2} \approx 10^{-13} \ \textrm{m}^{-1}$ to
$qB^{i}/mc \approx 10^{-11} \ \textrm{m}^{-1} $. We would like to add that the
maximum charge $q$ carried by a particle with a diameter $\Phi$ is
proportional to $\Phi^{2}$ \cite{Parker,Mizuno} while its mass $m$ is
proportional to $\Phi^{3}$. It follows that the ratio $q/m$ is a decreasing
function with respect to $q$ or $m$ since it is proportional to $\Phi^{-1}$.
This means that if we choose a test particle with a more great charge, its mass
would be great enough to obtain more small values for $\partial_{\nu}h^{0\mu}$.
In order to increase $\partial_{\nu}h^{0\mu}$, it is necessary to use a test
particle with a more small mass. Taking into account the value of the mass used
above, when $\partial_{\nu}h^{0\mu}$ reach significant values, the mass would
become so small that quantum effects would dominate. For example, if the test
particle is a proton, with the same values for $E$ and $B$ as above, we obtain
non negligible values, in the order of $qE^{i}/mc^{2} \approx 0.1 \ \textrm{m}^{-1}$
to $qB^{i}/mc \approx 1 \ \textrm{m}^{-1} $.

For non quantum systems, the fact that the derivatives $\partial_{\nu}h^{0\mu}$
are extremely close to zero means that $h^{0\mu}$ take practically constant
values. Since the metric must become Minkowskian for large distances, these
constants are close to zero. Thus, at every space time point, relations
(49) and (50) indicate clearly that even for strong electromagnetic field we
have $h^{0\mu} \ll 1$. This justify the linear approximation adopted here and
means that the higher order terms are negligible. In conclusion, Maxwell's
equations represent a first approximation of a new version of
Einstein's equations which describe electromagnetism.
Concerning the geodesic equation, as in section 2, using (41) instead of
(24), equation (32) reduces in lower velocity case to
\begin {equation}
m {d^2 x^{i} \over dt^2}
   =  q E^{i}+ q \left( \overrightarrow{v} \times \overrightarrow{B}\right)^{i} ,
\end {equation}
reproducing then the motion of a charged particle under Lorentz force.


\section{ Discussion}

Before to conclude, let us make two observations about these results.

The first concerns the Lorentz force law which is reproduced from the
geodesic equation without the time independence restriction for the
fields but in the lower velocity case. For arbitrary velocities, it is
easy to show that Lorentz force law never can be written as a geodesic
equation unless we consider the metric as a function of velocity.
This thwarts us in our wish to extend the PE. However, it is well known
that the Lorentz force law cannot be fundamental \cite{GHW,Jackson}
since it does not include self-force effects. One may wonder if the
complete law describing the electromagnetic force may be written as
a geodesic equation for arbitrary velocities. The question is left
opened and the possibility of extending the PE maybe depends on it.
The fact remains that we showed that Maxwell's equations are contained in
Einstein's ones regardless of the future of this extension of the PE.

The second observation concerns the non linear terms and the curvature
of space-time. If we take into account the higher-order terms, relation
(43) takes the form
\begin {equation}
G^{0\nu} = {q \over 2 m c } \partial_{\mu} F^{\mu \nu}
           + {q^{2} \over m^{2} }f^{(2)} + {q^{3} \over m^{3} }f^{(3)}+ \ldots ,
\end {equation}
where $f^{n}$ $(n=2,3, \ldots)$ is a function of $n$th order in $A$, its derivatives
and an analogous field defined with $h^{ij}$. Einstein's equations, $G^{0\nu} =0$,
yield
\begin {equation}
 {1 \over 2c } \partial_{\mu} F^{\mu \nu}
           + {q \over m }f^{(2)} + {q^{2} \over m^{2} }f^{(3)}+ \ldots =0.
\end {equation}
In gravity, the ratio $q/m $ must be substituted by $m_{g}/m_{i}$ which
keeps a constant value for any particle.
Thus, the gravitational analogous of (53) indicates that even in
the absence of a test particle, the higher-order terms remain. However,
in the electromagnetic case, the ratio $q/m$ does not take the same value
for various particles. In particular, if we choose a test particle such
that $q \rightarrow 0$ and $m_{i}\neq 0$, all the higher-order terms
in (53) vanish. In this case (53) reduces to the usual Maxwell equations,
$\partial_{\mu} F^{\mu \nu}=0$, and the space-time remains Minkowskian.
This feature can be seen also through the spherical symmetry case with the
use of the Schwarzschild solution. In fact, let us rewrite expression (37)
for an arbitrary charged test particle $q$ under the action of another
fixed charge $Q$
\begin {equation}
\xi_{Q}(r)=1-{2 K \over c^2 }{ Q \over r}{q \over m_{i} }.
\end {equation}
If $q \rightarrow 0$ and $m_{i}\neq 0$, Eq. (54) indicates that
$\xi_{Q}(r) \rightarrow 1$ and therefore the space-time
is Minkowskian. This means that the presence of the charge $Q$ is not
sufficient to affect the space-time geometry. Consequently, it is only in
the presence of an interaction between the source and a test particle that
the high-orders terms appear and that the space-time geometry is affected.
Maybe this is the fundamental distinction between the gravitational and the
electromagnetic interactions. This feature allows to pave the way for a new
concept of the field. As shown in the last section, we notice that the
effects of these higher-orders terms are negligible in the present domain
of high voltage. However, there effects become significant in the subatomic
physics where the test particle masses are very small and the quantum
effects dominate. This should explain why the Barros approach allows to
reproduce correctly the hydrogen atom spectrum.

We would like to add that this intriguing feature concerning the dependence
of the fields upon the test particle properties is not so strange.
In fact, in the Deformed Special Relativity, the coordinate
transformation law \cite{KMM} depends on the impulsion and energy
of the test particle. This will induce a dependence on the test particle
properties of the electromagnetic field in the high energy
domain \cite{Harikumar}

In conclusion, the present investigation can be summarized in two main
results. The first concerns the linearized gravity which we have revised
in such a way as to eliminate some imperfections appearing in the
earlier versions. In particular, we have introduced a subtle gauge so as
to get to a strong similarity with electromagnetism. In the context of
this result, it is interesting to reconsider the effects of the
magnetic-type gravitation. The second main result is based on this
similarity and on the Barros results and consists in extracting the Maxwell
equations from a new version of Einstein's ones. As the gravitational field,
it is possible to describe the effects of the electromagnetic interaction
directly in the metric without introducing a term describing the
interaction. Thus, Einstein's equations can be reinterpreted since they
can be adapted to describe other interactions, not only the gravitational
one.

\bigskip
\bigskip

\noindent
{\bf REFERENCES}
\vskip\baselineskip


\begin{enumerate}

\bibitem{Einstein1}
Einstein, A.: Session Report of the Prussian Academy of Sciences,
pp 414-419 (1925)

\bibitem{Einstein2}
Einstein, A.: arXiv e-print: physics/0503046

\bibitem{OW}
Overduin, J.M.,  Wesson,  P.S.: Phys. Rept.  {\bf 283}, 303-380 (1997), arXiv e-print: gr-qc/9805018

\bibitem{Huei}
Huei, P.: Gen. Rel. Grav. {\bf 15},  725-735 (1983)

\bibitem{Wald}
Wald, R.M.: General Relativity, Chap. 4, The University of Chicago Press,
Chicago (1984)

\bibitem{Carroll}
Carroll, S.: Spacetime and Geometry, Chap. 7, Addison Wesley, San Francisco (2004)

\bibitem{Barros1}
Barros, C.C.Jr.: Eur. Phys. J. {\bf C 42},  119-126 (2005), arXiv e-print: physics/0409064

\bibitem{Barros2}
Barros, C.C.Jr.: Eur. Phys. J. {\bf C 45}, 421-425 (2006), arXiv e-print: hep-ph/0504179

\bibitem{Barros3}
Barros, C.C.Jr.: arXiv e-print: Physics/0509011


\bibitem{Belabbas}
Belabbas, A.: Les potentiels non Gravitationnels et la Structure de l'Espace-temps, Magister Thesis,
University of B\'{e}jaia (2006), arXiv e-print: 0906.2519

\bibitem{Weinberg}
Weinberg, S.: Gravitation and Cosmology, John Wiley - Sons, Inc., U.S.A (1972)

\bibitem{Parker}
Parker, K.: Electrical Operation of Electrostatic Precipitators, Institution of Engineering
and Technology, London (2003)

\bibitem{Mizuno}
Mizuno, A.: IEEE Transactions on Dielectrics and Electrical Insulation {\bf 7}, 615-624 (2000)

\bibitem{GHW}
Gralla, S.E., Harte, A.I., Wald, R.M.:  arXiv e-print: 0905.2391

\bibitem{Jackson}
Jackson, J.D.: Classical Electrodynamic, Chap. 17, John Wiley - Sons, Inc.,
New York (1962)

\bibitem{KMM}
Kimberly, D., Magueijo, J.,  Medeiros, J.: Phys. Rev. {\bf D70} 084007 (2004), arXiv e-print: gr-qc/0303067

\bibitem{Harikumar}
Harikumar, E.:  arXiv e-print: 1002.3202


\end{enumerate}
\end {document}